\documentclass[journal]{IEEEtran}
\usepackage[hidelinks]{hyperref}
\usepackage{amsmath,amssymb,amsfonts,mathtools}
\usepackage{algorithmic}
\usepackage{amsthm}

\usepackage{graphicx}

\usepackage{textcomp}
\usepackage{nomencl}
\makenomenclature
\usepackage{etoolbox}
\renewcommand\nomgroup[1]{%
	\item[%
	\ifstrequal{#1}{A}{\textit{\underline{Sets}}}{%
		\ifstrequal{#1}{B}{\textit{\underline{Parameters}}}{%
			\ifstrequal{#1}{C}{\textit{\underline{Variables}}}{
				\ifstrequal{#1}{D}{\textit{\underline{Envelopes related notation}}}{
                       \ifstrequal{#1}{E}{\textit{\underline{Other operators}}}}}}}%
	]}

\PassOptionsToPackage{hyphens}{url}\usepackage{hyperref}

\usepackage{makecell}
\usepackage{textcomp}
\usepackage[noadjust]{cite}
\usepackage{multicol}
\usepackage{url}
\usepackage{mathrsfs} %% different font
\usepackage{booktabs} %% for professional tables
\usepackage{makecell} %% for spanning rows with text in tables
\usepackage[flushleft]{threeparttable} % for footnotes in tables

\usepackage{booktabs} %% for tables
\usepackage{makecell} %% two column cells
\usepackage{threeparttable}

\usepackage{eufrak} %% for differnt fonts

\usepackage{amsbsy} %% for boldface \mathscr

\usepackage{todonotes} %% for commenting

\usepackage[caption=false, font=footnotesize]{subfig}

\usepackage{circledsteps}

\usepackage{todonotes}

\usepackage{scalerel}
\usepackage{tikz}
\usetikzlibrary{svg.path}

\definecolor{orcidlogocol}{HTML}{A6CE39}
\tikzset{
	orcidlogo/.pic={
		\fill[orcidlogocol] svg{M256,128c0,70.7-57.3,128-128,128C57.3,256,0,198.7,0,128C0,57.3,57.3,0,128,0C198.7,0,256,57.3,256,128z};
		\fill[white] svg{M86.3,186.2H70.9V79.1h15.4v48.4V186.2z}
		svg{M108.9,79.1h41.6c39.6,0,57,28.3,57,53.6c0,27.5-21.5,53.6-56.8,53.6h-41.8V79.1z M124.3,172.4h24.5c34.9,0,42.9-26.5,42.9-39.7c0-21.5-13.7-39.7-43.7-39.7h-23.7V172.4z}
		svg{M88.7,56.8c0,5.5-4.5,10.1-10.1,10.1c-5.6,0-10.1-4.6-10.1-10.1c0-5.6,4.5-10.1,10.1-10.1C84.2,46.7,88.7,51.3,88.7,56.8z};
	}
}

\newcommand\orcidicon[1]{\href{https://orcid.org/#1}{\mbox{\scalerel*{
				\begin{tikzpicture}[yscale=-1,transform shape]
					\pic{orcidlogo};
				\end{tikzpicture}
			}{|}}}}

\usepackage{titlesec} 

\usepackage[linesnumbered, ruled]{algorithm2e}
\SetKwRepeat{Do}{do}{while} %% for do-while loop

\newcommand{\userAll}{\mathcal{H}}
\newcommand{\DOEuser}{\ddot{\mathcal{H}}}
\newcommand{\PVuser}{\dot{\mathcal{H}}}
\newcommand{\passuser}{\hat{\mathcal{H}}}

\newcommand{\Time}{\mathcal{T}} %% for time 
\newcommand{\Node}{\mathcal{N}} %% for nodes of the network

\theoremstyle{remark}
\newtheorem*{remark}{Remark}

\theoremstyle{definition}

\begin{document}

\title{Development and Validation of a Dynamic Operating Envelopes-enabled Demand Response Scheme in Low-voltage Distribution Networks}

% \title{Software-in-the-loop Development of Demand Response in Low-voltage Networks under the Dynamic Operating Envelopes Framework}
	
\author{Gayan~Lankeshwara \orcidicon{0000-0002-1899-4571}\,,~\IEEEmembership{Member,~IEEE,}
	Rahul~Sharma~\orcidicon{0000-0002-2440-5286}\,,~\IEEEmembership{Senior Member,~IEEE,}
 	M.~R.~Alam~\orcidicon{0000-0001-6017-764X}\,,~\IEEEmembership{Member,~IEEE,}
	Ruifeng~Yan~\orcidicon{0000-0002-5779-9090}\,,~\IEEEmembership{Member,~IEEE,} and
	Tapan~K. Saha \orcidicon{0000-0003-0763-0032}\,,~\IEEEmembership{Fellow,~IEEE}% <-% <-this % stops a space
	\thanks{Gayan Lankeshwara, Rahul Sharma, M. R. Alam, Ruifeng Yan and Tapan K. Saha are with the School of Electrical Engineering and Computer Science, The University of Queensland, Brisbane, QLD 4072, Australia (email: g.lankeshwara@uq.net.au, rahul.sharma@uq.edu.au, alam.mollah9@gmail.com, ruifeng@eecs.uq.edu.au and saha@eecs.uq.edu.au)}}% <-this % stops a space

\markboth{submitted to IEEE Transactions on Power Systems}%
{Shell \MakeLowercase{\textit{et al.}}: Bare Demo of IEEEtran.cls for IEEE Journals}
	
\maketitle
	
\begin{abstract}
Dynamic operating envelopes (DOEs) offer an attractive solution for  maintaining network integrity amidst increasing penetration of distributed energy resources (DERs) in low-voltage (LV) networks. Currently, the focus of DOEs primarily revolves around active power exports of rooftop photovoltaic (PV) generation, often neglecting the impact of demand response (DR). This paper presents a two-stage, coordinated approach for residential DR participation in electricity markets under the DOE framework. In the first stage, the distribution network service provider (DNSP) adopts a \textit{convex hull} technique to establish DOEs at each customer point-of-connection (POC). In the second stage, the demand response aggregator (DRA) utilises DOEs assigned by the DNSP to develop a hierarchical control scheme for tracking a load set-point signal without jeopardising network statutory limits.
To assess the effectiveness of the proposed control scheme in a practical setting, software-in-the-loop (SIL) tests are performed in a grid simulator, considering a real residential feeder with realistic household load and generation profiles. Simulation validations suggest that the DRA can provide precise DR while honouring network statutory limits and maintaining end-user thermal comfort. Furthermore, the overall approach is compliant with the market dispatch interval and preserves end-user data privacy. 
\end{abstract}


\section{Introduction}
\label{sec: Introduction}
The proliferation of behind-the-meter distributed energy resources (DERs) in low-voltage (LV) networks and their participation in demand response (DR) services have escalated technical challenges faced by distribution network service providers (DNSPs) in maintaining secure operation of the network. The current strategy to overcome these challenges is to impose fixed import/export limits at the customer point-of-connection (POC) and curtail excess consumption/generation beyond the predefined limit. For instance, one of the DNSPs in Queensland, assigns 5 kW export limits for small customer connections \cite{Energex_standards_for_small_connections}. However, as static limits are calculated based on rarely occurring load and generation scenarios, they often underutilise the available capacity of distributed energy resources (DERs).
\par
\textit{Dynamic operating envelopes} (DOEs) is an insightful approach that takes account of the dynamic behaviour of household load and generation in determining import/export limits \cite{CellPress_Tushar}. According to \cite{dyanmic_envelopes_outcome_report}, DOEs are operating envelopes that vary import and export limits over time and location based on the available capacity of the network. Aligned with this, DOEs can be implemented at either DER-level or at the POC of an end-user \cite{Blackhall2020}.
\par
The strategies for implementing DOEs for households can be mainly discussed under two groups: 1) optimal power flow (OPF)-based allocation; 2) network sensitivity-based allocation. In OPF-based allocation method, an optimisation problem is formulated at the DNSP layer with the objective of maximising utility and/or social welfare while taking into account network configuration and parameters, statutory limits, operational and technical limits for individual DER assets. The solution provides optimal power set-points for customer connections or individual DERs such that network technical limits are not jeopardised. Following this, the authors in \cite{Nando_OPF_paper} have proposed a three-phase, centralised OPF scheme for a DNSP to calculate operating envelopes in terms of import and export limits at meter-level while ensuring network integrity. In \cite{ensuring_petrou}, the authors have presented a rule-based approach where DOEs are imposed for household connections only if the intended operation compromises network voltage limits. Furthermore, only active power injections at the POC are considered. In addition to that, there is no guarantee that end-user data privacy \cite{NIST_privacy} is preserved as the DNSP requires forecasts of household load and generation to determine DOEs. To address end-user privacy violations, the authors in \cite{Attarha_network_secure_envelopes} have proposed a hierarchical control scheme based on the alternating direction method of multipliers (ADMM) where DOEs are established to control rooftop photovoltaic (PV) and battery storage to provide services in electricity markets. Moreover, this approach also accommodates end-user reactive power through a Q-P controller.
\par
Ref. \cite{Techno_economic_DOE} proposes a two-stage, top-down approach to allocate DOEs in medium voltage (MV)-LV integrated networks. In the first stage, state estimation and capacity constrained optimisation is utilised to allocate DOEs at transformer-level. In the next stage, based on transformer-level envelopes, a three-phase unbalanced AC OPF scheme is adopted to allocate DOEs for household connections in LV networks. In \cite{DOE_with_flexibility}, the authors have developed a near real-time approach to determine DOEs that explicitly account for end-user flexibility in LV distribution networks. However, the proposed scheme does not provide any information on how established DOEs can be utilised to provide network-aware DR services.
% In \cite{DOE_driven_P2P}, a network-aware peer-to-peer (P2P) trading scheme is presented for LV distribution networks where DOEs in terms of active power injections at household POCs are calculated by solving an AC OPF problem.
\par
In network sensitivity-based allocation of DOEs, the sensitivity of voltage and current with respect to active and reactive power are determined for operating state of the network via different methods. Thereafter, sensitivity factors are used to determine nodal injections of active and reactive power such that network constraints are not jeopardised. For instance, the authors in \cite{Rigoni_UCD} have presented a coordinated DR scheme where the DNSP calculates network sensitivities at the POC using a linear regression technique and submits to the demand response aggregator (DRA) for its day-ahead scheduling and real-time operation in electricity markets. Furthermore, the overall approach ensures safe operation within network statutory limits.
% In \cite{Rahul_Gupta}, a network-aware method based on distributed model predictive control (MPC) and ADMM is utilised for DERs to track a day-ahead dispatch plan and for the operation in real-time. Furthermore, an analytical approach is occupied to determine network sensitivities.
\par
To this end, most of the existing literature has focussed on 
dynamic operating envelopes for export power management--cater for the strong uptake of rooftop PV and battery storage in LV distribution networks \cite{Nando_OPF_paper, ensuring_petrou, Attarha_network_secure_envelopes, Techno_economic_DOE, DOE_with_flexibility}. However, with end-customers actively participating in wholesale DR schemes \cite{WholesaleDR} and the increasing popularity of electric vehicles \cite{EV_stats_Australia}, it is undoubted that network violations would continue to exacerbate on top of households exporting rooftop PV generation to the grid \cite{Project_EDGE_report}. In that respect, only a handful of existing studies have focussed on network-aware residential DR under the DOE framework. Despite the consideration of residential DR in \cite{Rigoni_UCD}, the overall centralised approach requires end-users to share sensitive information with the DNSP and the DRA. This gives rise to end-user data privacy concerns \cite{NIST_privacy}. On the other hand, dynamic import limits will also be important when residential DR is present. This has been overlooked in the DOE outcomes report published by Australian Renewable Energy Agency \cite{dyanmic_envelopes_outcome_report}. Furthermore, the adaptability of a DOE-driven DR scheme in a practical setting heavily relies on the coordination between stakeholders, especially the privacy and separation between the DNSP and the DRA. Although a coordinated DR scheme under the DOE framework is proposed in \cite{lankeshwara_dynamic_2022}, the overall approach lacks real-time validations. Furthermore, the computation time requirements are not explicitly mentioned. To this end, it is equally important to develop scalable architectures such that the assignment of DOEs and real-time DR control action are performed within the dispatch interval of the market operator.
% \begin{itemize}
%     \item Although a DOE scheme that captures end-user flexibility is introduced in \cite{DOE_with_flexibility}, the approach is limited to PV power exports and does discuss the engagement of demand response and its effects. Moreover, a real-time control scheme is also not proposed.
%     \item Monash paper
%     \item Brian Liu and Gregor Verbic IEEE paper
% \end{itemize}
% }
\par 
The main contributions of this work can be summarised as follows:
\begin{itemize}
		\item Developing a two-stage, coordinated control scheme for dynamic operating envelopes-enabled demand response in low-voltage networks.
        \item A software-in-the-loop validation of the proposed control scheme in the grid simulator to assess the scalability in a practical setting.
        \item Dynamic operating envelopes outline both active and reactive, import and export power limits at household connection point. This information is useful for the demand response aggregator in the market bidding process.
		\item Preserving end-user data privacy for the real-time demand response via a hierarchical implementation based on the ADMM form of the resource-sharing problem.

\end{itemize}

% The rest of the paper is organised as follows. Section~\ref{sec: Proposed_methodology} describes the proposed methodology. The simulation studies are discussed in section \ref{sec: Results} and concluding remarks are presented in section \ref{sec: Conclusions}.
\begin{figure}[t]
	\centering
    \includegraphics[trim={0 10pt 0 5pt}, clip,width=\columnwidth]{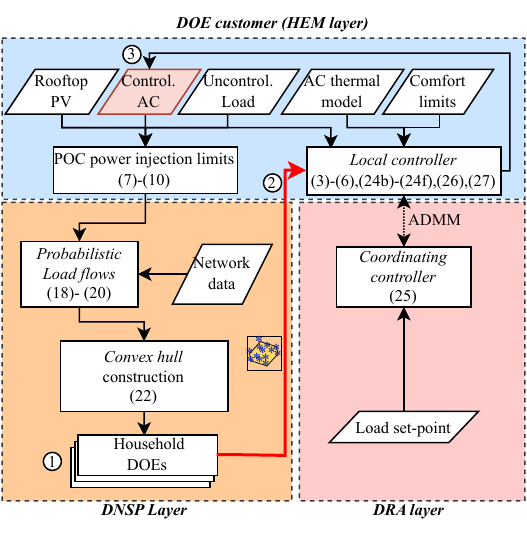}
	\caption{A summarised block diagram of the proposed approach (corresponding equation numbers are included in brackets)}
	\label{fig:summarised_block_diagram}
\end{figure}
\section{Proposed Methodology}
\label{sec: Proposed_methodology}
In this work, a two-stage coordinated approach is proposed for the DOE-enabled DR in residential LV networks. A summarised block diagram of the proposed approach is given in Fig. \ref{fig:summarised_block_diagram}. According to the figure, the control architecture is discussed under three layers, home energy management (HEM) layer for end-users, DNSP layer and the DRA layer.  Following this, the overall control operation can be described under two stages. In the first stage, the DNSP determines household DOEs based on power injection limits at POC (\Circled{1} in Fig.~\ref{fig:summarised_block_diagram}) and shares with the DRA (\Circled{2} in Fig.~\ref{fig:summarised_block_diagram}). In the next stage, the DRA utilises DOEs determined by the DNSP to provide DR services while concurrently satisfying network statutory limits (\Circled{3} in Fig.~\ref{fig:summarised_block_diagram}). 
\par
Let $\userAll:=\{1,2,\ldots,H\}$ be the set of households indexed by $h$ and  $\Time:=\{t_{0},t_{1},\ldots,t_{0}+T\}$ be the set of time periods indexed by $t$. As per existing regulatory standards in Australia, only end-users equipped with embedded generation are allowed to adopt dynamic connections \cite{Energex_standards_for_small_connections}. Following this, three disjoint sets of residential customers are defined. The first group is  \emph{DOE customers} ($\DOEuser \subseteq \userAll$), who are equipped with rooftop PV generation and adopt DOEs. Moreover, \emph{DOE customers} will only participate in DR events. The next group is \textit{non-DOE customers} ($\PVuser \subseteq \userAll$), who are equipped with rooftop PV but do not adopt DOEs. The final group is \textit{passive customers} ($\passuser \subseteq \userAll$). This group consists of end-users who are not equipped with rooftop PV generation. The relationship between each customer categories can be mathematically expressed as: $\DOEuser \cup \PVuser \cup \passuser = \userAll$. Following this, the two stage approach of DOE-enabled DR is discussed in detail.

\subsection{DNSP determining household operating envelopes (Stage~I)}
\label{subsec: DNSP determining operating envelopes in real-time}
The process of DNSP establishing household operating envelopes consists of two steps. In the first step, households calculate active and reactive power injection limits at the POC for a certain time step. In the next step, the DNSP constructs household operating envelopes based on a combination of probabilistic load flows and a \textit{convex hull} approximation. 

\subsubsection{Determining household power injection limits}
\label{subsubsec: Households determining power injection limits}
This is the first step of the overall process of establishing DOEs at the POC of households. Let us consider inverter-type air-conditioners to be the controllable DER asset for \textit{DOE customers} participating in DR. Therefore, active and reactive power injection at the POC of household $h\in\DOEuser$ at time $t\in\Time$ can be expressed as: 
\begin{align}
	\label{eq:active-power-injection}
	P_{\mathrm{inj},t}^{\,h}&= \tilde{P}_{\mathrm{PV},t}^{\,h}-P_{\mathrm{AC},t}^{\,h}-\tilde{P}_{\mathrm{UL},t}^{\,h}\\
	\label{eq:reactive-power-injection}
	Q_{\mathrm{inj},t}^{\,h}&= \tilde{Q}_{\mathrm{PV},t}^{\,h}-Q_{\mathrm{AC},t}^{\,h}-\tilde{Q}_{\mathrm{UL},t}^{\,h}
\end{align}
where $P_{\mathrm{inj},t}^{\,h}$ and $Q_{\mathrm{inj},t}^{\,h}$ corresponds to active and reactive power injection at the POC,  $\tilde{P}_{\mathrm{PV},t}^{\,h}$ and $\tilde{Q}_{\mathrm{PV},t}^{\,h}$ correspond to available active and reactive power generation of rooftop PV, $P_{\mathrm{AC},t}^{\,h}$, $Q_{\mathrm{AC},t}^{\,h}$ correspond to active and reactive power consumption of the air-conditioner, $\tilde{P}_{\mathrm{UL},t}^{\,h}$ and $\tilde{Q}_{\mathrm{UL},t}^{\,h}$ correspond to active and reactive power consumption of uncontrollable loads of household $h$ at time $t$. Aligned with the sign convention, $P_{\mathrm{inj},t}^{\,h}>0$ represents exporting power to the grid whereas $P_{\mathrm{inj},t}^{\,h}<0$ represents importing power from the grid at time $t$. 
\par
For the controllable air-conditioning load, active power is bounded by,
\begin{equation}
	\label{eq:active-power-bounds-AC}
	0 \leq P_{\mathrm{AC},t}^{\,h} \leq \overline{P}_{\mathrm{AC}}^{\,h}
\end{equation}
where $\overline{P}_{\mathrm{AC}}^{\,h}$ is the rated power of the inverter-type air-conditioner at house $h\in\DOEuser$. It is important highlighting that, inverter-type air-conditioner can operate under continuous power consumption levels as compared to  regular ON-OFF type air-conditioners where the power consumption is only limited to zero and rated power \cite{MahdaviPaper}.
Furthermore, it is assumed that the active power factor correction method \cite{active_power_factor_correction} is employed for the operation of the inverter-type air-conditioner. Hence, the relationship between $P_{\mathrm{AC},t}^{\,h}$ and $Q_{\mathrm{AC},t}^{\,h}$ at time $t$ can be expressed as:
\begin{equation}
	\label{eq: air-conditioner-power-relationship}
	{Q}_{\mathrm{AC},t}^{\,h} = {P}_{\mathrm{AC},t}^{\,h} \cdot \tan\left(\cos^{-1}\left(\varphi_{{\mathrm{AC}}}^{\,h}\right)\right)
\end{equation}
where $\varphi_{{\mathrm{AC}}}^{\,h}$ is the constant power factor (lagging) for the inverter-type air-conditioner at household $h\in\DOEuser$.
\par
Assuming rooftop PV generation to be uncontrollable, fixed power factor operation in line with \textit{AS/NZS 4777.2} standards \cite{AS4777_standards} is assumed. Hence, $\tilde{Q}_{\mathrm{PV},t}$ of rooftop PV in house $h\in\DOEuser$ can be obtained as:
\begin{equation}
	\label{eq: pv-inverter-power-relationship}
	\tilde{Q}_{\mathrm{PV},t}^{\,h} = \tilde{P}_{\mathrm{PV},t}^{\,h} \cdot \tan\left(\cos^{-1}\left(\varphi_{{\mathrm{PV}}}^{\,h}\right)\right)
\end{equation}
where $\varphi_{{\mathrm{PV}}}^{\,h}$ corresponds to the power factor of the rooftop PV inverter at household $h\in\DOEuser$.
\par
Similarly, for uncontrollable loads, the operation under fixed power factor is assumed. Following this, the relationship between active and reactive power can be expressed as:
\begin{equation}
	\label{eq:uncontrollable-load-power-relationship}
	\tilde{Q}_{\mathrm{UL},t}^{\,h} = \tilde{P}_{\mathrm{UL},t}^{\,h} \cdot \tan\left(\cos^{-1}\left(\varphi_{{\mathrm{UL}}}^{\,h}\right)\right)
\end{equation}
where $\varphi_{{\mathrm{UL}}}^{\,h}$ corresponds to the power factor of the uncontrollable load at household $h\in\DOEuser$.
\par
Consequently, the minimum and maximum limits of active and reactive power injection at the POC of household $h\in\DOEuser$ at time $t$ can be algebraically calculated as:
\begin{align}
	\label{eq:minimum-active-power-limits}
	\underline{P}_{\,\mathrm{inj},t}^{\,h} &= \min \left(	P_{\mathrm{inj},t}^{\,h}\right)\quad \text{s.t. }
	\eqref{eq:active-power-injection}-\eqref{eq:uncontrollable-load-power-relationship} \\
	\label{eq:maximum-active-power-limits}
	\overline{P}_{\,\mathrm{inj},t}^{\,h} &= \max \left(	P_{\mathrm{inj},t}^{\,h}\right)\quad \text{s.t. } \eqref{eq:active-power-injection}-\eqref{eq:uncontrollable-load-power-relationship} \\
	\label{eq:minimum-reactive-power-limits}
	\underline{Q}_{\,\mathrm{inj},t}^{\,h} &= \min \left(	Q_{\mathrm{inj},t}^{\,h}\right)\quad \text{s.t. } \eqref{eq:active-power-injection}-\eqref{eq:uncontrollable-load-power-relationship} \\
	\label{eq:maximum-reactive-power-limits}	
	\overline{Q}_{\,\mathrm{inj},t}^{\,h} &= \max \left(Q_{\mathrm{inj},t}^{\,h}\right)\quad \text{s.t. } \eqref{eq:active-power-injection}-\eqref{eq:uncontrollable-load-power-relationship}
\end{align}
For instance, the minimum active power injection for household $h\in\DOEuser$ at time $t$, $\underline{P}_{\,\mathrm{inj},t}^{\,h}$ can be calculated by considering the maximum of $P_{\mathrm{AC},t}^{\,h}$ which is $\overline{P}_{\mathrm{AC}}^{\,h}$ according to \eqref{eq:active-power-bounds-AC}. This is because the air-conditioner is the controllable load in DR and the active power injection at the POC varies according to the value of $P_{\mathrm{AC},t}^{\,h}$. In a similar way, the maximum active power injection, $\overline{P}_{\,\mathrm{inj},t}^{\,h}$, in \eqref{eq:maximum-active-power-limits} can be calculated by considering the minimum of  $P_{\mathrm{AC},t}^{\,h}$ which is zero.
\par
For \textit{non-DOE customers}, i.e., $h\in\PVuser$ who do not participate in DR events, the air-conditioning load is assumed to be uncontrollable and is captured in uncontrollable consumption. Hence,  \eqref{eq:active-power-injection} and \eqref{eq:reactive-power-injection} simplifies to: 
\begin{align}
	\label{eq:active-power-injection-non-DOE}
	P_{\mathrm{inj},t}^{\,h}&= \tilde{P}_{\mathrm{PV},t}^{\,h}-\tilde{P}_{\mathrm{UL},t}^{\,h}\\
	\label{eq:reactive-power-injection-non-DOE}
	Q_{\mathrm{inj},t}^{\,h}&= \tilde{Q}_{\mathrm{PV},t}^{\,h}-\tilde{Q}_{\mathrm{UL},t}^{\,h}
\end{align}
Since rooftop PV generation is also uncontrollable, active and reactive power limits at the POC simplifies to:
\begin{gather}
    \label{eq:active-power-injection-limits-non-DOE}    \underline{P}_{\,\mathrm{inj},t}^{\,h}=\overline{P}_{\,\mathrm{inj},t}^{\,h} = P_{\mathrm{inj},t}^{\,h}\\
    \label{eq:rective-power-injection-limits-non-DOE} Q_{\mathrm{inj},t}^{\,h}=\underline{Q}_{\,\mathrm{inj},t}^{\,h}=\overline{Q}_{\,\mathrm{inj},t}^{\,h}\\
    \text{subject to  \eqref{eq: pv-inverter-power-relationship} and \eqref{eq:uncontrollable-load-power-relationship}} \nonumber
    % \intertext{subject to  \eqref{eq: pv-inverter-power-relationship} and \eqref{eq:uncontrollable-load-power-relationship}} \nonumber
\end{gather}
For \textit{passive customers}, i.e., $h\in\passuser$, as air-conditioning load is uncontrollable and rooftop PV generation is absent, \eqref{eq:active-power-injection} and \eqref{eq:reactive-power-injection} simplifies to: 
\begin{align}
	\label{eq:active-power-injection-passive}
	P_{\mathrm{inj},t}^{\,h}&= -\tilde{P}_{\mathrm{UL},t}^{\,h}\\
	\label{eq:reactive-power-injection-passive}
	Q_{\mathrm{inj},t}^{\,h}&= -\tilde{Q}_{\mathrm{UL},t}^{\,h}
\end{align}
Similar to \textit{non-DOE customers}, active and reactive power injection limits at the POC for \textit{passive customers} can be expressed as \eqref{eq:active-power-injection-limits-non-DOE} and \eqref{eq:rective-power-injection-limits-non-DOE}. It should be noted minimum and maximum limits of active and reactive power injection coincide for both \textit{non-DOE customers} and \textit{passive customers}.
\par
Once $\left[\underline{P}_{\,\mathrm{inj},t}^{\,h},\overline{P}_{\,\mathrm{inj},t}^{\,h},\underline{Q}_{\,\mathrm{inj},t}^{\,h},\overline{Q}_{\,\mathrm{inj},t}^{\,h}\right]$ is calculated for all $h\in\userAll$ for the next time step, the information is passed to the DNSP. It should be hight that the calculation of minimum and maximum power injection limits could be easily embedded in existing HEM infrastructure \cite{HEMS}.

\subsubsection{Constructing household dynamic operating envelopes}
\label{subsubsec: Determining dynamic operating envelopes for households}
With information on power injection limits at the POC for all $h\in\userAll$ at a particular time step, the DNSP adopts a \textit{bounding-box approximation} method to determine the region of operation in the P-Q plane for each \textit{DOE-customer}. Considering household $h\in\DOEuser$ at time $t\in\Time$, the bounding box $\mathfrak{B}^{h}_{t}$ which captures the overall region of operation is given by:
\begin{equation}
	\label{eq:bouding box approaxmation}
	\mathfrak{B}^{h}_{t}:=\left\{\left(P_{\mathrm{inj},t}^{\,h},Q_{\mathrm{inj},t}^{\,h}\right)\;\middle\vert\;
	\begin{gathered}
		\underline{P}^{\,h}_{\mathrm{\,inj},t}\leq P_{\mathrm{inj},t}^{\,h} \leq \overline{P}^{\,h}_{\mathrm{\,inj},t} \\
		\underline{Q}^{\,h}_{\mathrm{\,inj},t}\leq Q_{\mathrm{inj},t}^{\,h} \leq \overline{Q}^{\,h}_{\mathrm{\,inj},t}
	\end{gathered}\right\}
\end{equation}
The bounding box in \eqref{eq:bouding box approaxmation} represents the capability curve at the POC for each \textit{DOE-customer} at time $t\in\Time$. Nonetheless, there is no guarantee that every combination of  $\left(P^{\,h}_{\mathrm{\,inj},t},Q^{\,h}_{\mathrm{\,inj},t}\right)\in \mathfrak{B}^{h}_{t}$ for $h\in\DOEuser$ would not jeopardise voltage statutory limits of the network. Therefore, the DNSP further utilises a probabilistic load flow approach to determine feasible pairs of $\left(P^{\,h}_{\mathrm{\,inj},t},Q^{\,h}_{\mathrm{\,inj},t}\right)\in \mathfrak{B}^{h}_{t}$ that would not not breach voltage limits at any node $i\in\Node$ of the network when a three-phase unbalanced load flows is performed. To this end, it is assumed that the DNSP has full information on configuration and parameters of the LV distribution network.
\par
Let $\Omega$ be the set of probabilistic load flow scenarios indexed by $\omega$ at time $t\in\Time$ 
and 
$\left(P_{\mathrm{inj},t}^{\,h,\omega},Q_{\mathrm{inj},t}^{\,h,\omega}\right)$ 
be the pair of active and reactive power injection at the POC of household 
$h\in\DOEuser$ under scenario $\omega$ at time $t$. For household $h\in\DOEuser$,  
$\left(P_{\mathrm{inj},t}^{\,h,\omega},Q_{\mathrm{inj},t}^{\,h,\omega}\right)$ 
are chosen from a uniform distribution such that  
$P_{\mathrm{inj},t}^{\,h,\omega}\in 
\mathcal{U}(\underline{P}^{\,h}_{\mathrm{\,inj},t},	
\overline{P}^{\,h}_{\mathrm{inj},t})$ and  $Q_{\mathrm{inj},t}^{\,h,\omega}\in 
\mathcal{U}(\underline{Q}^{\,h}_{\mathrm{\,inj},t},	
\overline{Q}^{\,h}_{\mathrm{inj}})$, where $\mathcal{U}(\cdot)$ represent a uniform distribution. As mentioned in section~\ref{subsubsec: Households determining power injection limits}, since 
$\left(P_{\mathrm{inj},t}^{\,h},Q_{\mathrm{inj},t}^{\,h}\right)$ is fixed for 
$h\in\PVuser\cup\passuser$, the bounding-box in \eqref{eq:bouding box 
approaxmation} simplifies to a single point in the P-Q plane. Hence,
$\left(P_{\mathrm{inj},t}^{\,h,\omega},Q_{\mathrm{inj},t}^{\,h,\omega}\right)$ 
is considered to be fixed for all $\omega\in\Omega$. Thereafter, a three-phase unbalanced load flow is performed to determine whether 
$\left(P_{\mathrm{inj},t}^{\,h,\omega},Q_{\mathrm{inj},t}^{\,h,\omega}\right)$ 
for $h\in\userAll$ is feasible. The three-phase unbalanced load flow equations are described by:
\begin{gather}
	\label{eq:three_phase_active_power_flow}
	\begin{gathered}
		P_{i,t}^{s} = \mathfrak{R}(V_{i,t}^{s})\sum_{k\in\mathcal{N}}\sum_{\gamma \in \Phi}\left[G_{ik}^{s\gamma}\cdot \mathfrak{R}(V_{k,t}^{s})-B_{ik}^{s\gamma}\cdot \mathfrak{I}(V_{k,t}^{s})\right]\\
		+\mathfrak{I}(V_{i,t}^{s})\sum_{k\in\mathcal{N}}\sum_{\gamma \in \Phi}\left[G_{ik}^{s\gamma}\cdot \mathfrak{I}(V_{k,t}^{s})+B_{ik}^{s\gamma}\cdot \mathfrak{R}(V_{k,t}^{s})\right]\\
		 i\in \mathcal{N}, s\in \Phi, t\in \mathcal{T}
	\end{gathered}\\
		\label{eq:three_phase_reactive_power_flow}
	\begin{gathered}
	Q_{i,t}^{s} = \mathfrak{I}(V_{i,t}^{s})\sum_{k\in\mathcal{N}}\sum_{\gamma \in \Phi}\left[G_{ik}^{s\gamma}\cdot \mathfrak{R}(V_{k,t}^{s})-B_{ik}^{s\gamma}\cdot \mathfrak{I}(V_{k,t}^{s})\right]\\
	-\mathfrak{R}(V_{i,t}^{s})\sum_{k\in\mathcal{N}}\sum_{\gamma \in \Phi}\left[G_{ik}^{s\gamma}\cdot \mathfrak{I}(V_{k,t}^{s})+B_{ik}^{s\gamma}\cdot \mathfrak{R}(V_{k,t}^{s})\right]\\
	i\in \mathcal{N}, s\in \Phi, t\in \mathcal{T}
\end{gathered}
\end{gather}
where $\mathfrak{R}(\cdot)$ and $\mathfrak{I}(\cdot)$ represents \emph{real} and \emph{imaginary} parts of a complex number; $\mathcal{N}$ is the set of three-phase P-Q buses indexed by $i, k$; $\Phi$ is the phases indexed by $s,\gamma$; $G_{ik}^{s\gamma}$ is the conductance and $B_{ik}^{s\gamma}$ is the susceptance between phase $s$ and $\gamma$ for the line between bus $i$ and $k$ and $V_{i,t}^{s}$ is the complex voltage of phase $s$ in bus $i$ at time $t$; $P_{i,t}^{s}$ and $Q_{i,t}^{s}$ represents active and reactive power injection at phase $s$ in bus $i$ at time $t$. Assuming each phase of a particular bus has a single household connected to it, $P_{i,t}^{s}$ and $Q_{i,t}^{s}$ can be equivalently represented as $P_{\mathrm{inj},t}^{\,h}$ and $Q_{\mathrm{inj},t}^{\,h}$. 
\par
The voltages of the resultant load flow should remain within technical limits described by:
\begin{equation}
	\label{eq:voltage-statutory-limits}
	\underline{v}\leq \lvert V^{s}_{i,t} \rvert \leq \overline{v}, \quad i\in\Node, \,s\in \phi
\end{equation}   
where $\underline{v}$ and $\overline{v}$ are the lower and upper limits of voltage.
\par
Following this, the feasible region of operation in the P-Q plane for $h\in\DOEuser$ at time $t$, $\mathcal{B}^{h}_{t}\subseteq \mathfrak{B}^{h}_{t}\subset \mathbb{R}^2$, can be expressed as:
\begin{equation}
	\label{eq: feasbile region without voltage violations}
	\mathcal{B}^{h}_{t}:=\left\{\left(P_{\mathrm{inj},t}^{\,h,\omega},Q_{\mathrm{inj},t}^{h, \omega}\right), \omega \in \Omega\:\middle\vert\;
	\begin{gathered}
			\textrm{satisfies \eqref{eq:voltage-statutory-limits},}\\
			\textrm{\eqref{eq:three_phase_active_power_flow}, \eqref{eq:three_phase_reactive_power_flow}} 
	\end{gathered}\right\}
\end{equation}   
Compared to a set of feasible pairs of $\left(P_{\mathrm{inj},t}^{\,h},Q_{\mathrm{inj},t}^{\,h}\right)$ that would not violate voltage constraints at any node of the network at a particular time step, the DNSP would be interested in establishing an envelope which outlines the overall feasible region of operation at the POC for all $h\in\DOEuser$. To this end, the \textit{convex hull} \cite{de2000computational} of $\left(P_{\mathrm{inj},t}^{\,h},Q_{\mathrm{inj},t}^{\,h}\right)$ is obtained as the envelope at the POC.
\par
Let $\left[P_{\mathrm{inj},t}^{\,h,\omega},Q_{\mathrm{inj},t}^{h, \omega}\right]^{\mathrm{T}}\in \mathbb{R}^{2}$ be a feasible pair of power injection for house $h\in\DOEuser$ under scenario $\omega$ at time $t$, then the \textit{convex hull} of all the feasible pairs of power injections for $h\in\DOEuser$ at time $t\in\Time$, i.e., the \textit{convex hull} of $\mathcal{B}^{h}_{t}$, can be obtained as:
\begin{equation}
	\label{eq:convex-hull-for-households}
	\mathbf{conv}(\mathcal{B}^{h}_{t}) := \left\{\sum_{\omega}\theta_{\omega}\,\left[P_{\mathrm{inj},t}^{\,h,\omega},Q_{\mathrm{inj},t}^{h, \omega}\right]^{\mathrm{T}}\; \middle \vert\;
	\begin{gathered}
		\left[P_{\mathrm{inj},t}^{\,h,\omega},Q_{\mathrm{inj},t}^{h, \omega}\right]^{\mathrm{T}}\in \mathcal{B}^{h}_{t} \\
		\theta_{\omega}\geq 0 \textrm{ for all } \omega \\
		 \sum_{\omega}\theta_{\omega} =1
	\end{gathered}\right\}
\end{equation}
It is important highlighting that, considering the 2-D nature of the problem, the convex envelope defined in \eqref{eq:convex-hull-for-households} reduces to a convex polygon such that $\left(P_{\mathrm{inj},t}^{\,h,\omega},Q_{\mathrm{inj},t}^{h, \omega}\right)\in\mathbb{R}^{2}$. Moreover, considering the \textit{half-space representation} \cite{de2000computational} for a convex polygon, $\mathbf{conv}(\mathcal{B}^{h}_{t})$ can be expressed as:
\begin{equation}
	\label{eq:operating_envelopes_H_representation}
	\mathbf{conv}(\mathcal{B}^{h}_{t}) = \left\{\left(P_{\mathrm{inj},t}^{h},Q_{\mathrm{inj},t}^{h}\right)\Big|\mathbf{A}\cdot \left[P_{\mathrm{inj},t}^{\,h},Q_{\mathrm{inj},t}^{h}\right]^{\mathrm{T}}\leq \mathbf{b}\right\}
\end{equation}
where $\mathbf{A}\in \mathbb{R}^{m\times 2}$ and $\mathbf{b}\in\mathbb{R}^{m}$ such that $m\leq \vert \mathcal{B}^{h}_{t} \vert$, $\vert \cdot \vert$ represents the cardinality of a set. To this end, the \textit{convex hull} and parameters of its half-space representation can be calculated using existing software packages such as MATLAB and MPT-3 \cite{MPT3_toolbox}. Once operating envelopes are calculated for a certain time step, they are shared with \textit{DOE customers} (as in Fig.~\ref{fig:summarised_block_diagram}) for their participation in DR events.
\par
Unlike DOEs obtained as explicit set-points determined based on certain design objectives as in OPF approaches \cite{Nando_OPF_paper, ensuring_petrou, Attarha_network_secure_envelopes}, the DOEs in the proposed approach outlines the feasible region of operation in the P-Q plane, i.e., active-reactive and import-export limits without breaching voltage limits of the network. 

\subsection{DRA controlling household air-conditioners to provide DR services (Stage II) }
\label{subsec: DRA controlling household loads in real-time}
Once DOEs are shared by DNSP for customer connections, the DRA provides DR in real-time by controlling the consumption of air-conditioning loads belonging to \textit{DOE customers}. Aligned with this, the objective of each \textit{DOE customer} is to maximise financial returns by exporting power to the grid while maintaining indoor temperature within thermal comfort limits. In addition to that, the operation at the POC is constrained by the DOEs shared by the DNSP. 
\par
Considering $h\in\DOEuser$ at time $t\in\Time$, the optimisation problem can be expressed as:
\begin{subequations}
	\label{eq:household_optimisation_problem}
\begin{gather}
	\label{eq:household_obejctive_function}
	\underset{P_{\mathrm{AC}}^{\,h}}{\mathrm{max}}\quad \pi_{t}^{h}\cdot P_{\mathrm{inj},t}^{\,h}\\
	\shortintertext{subject to:}
	\label{eq:local-opt-active-power-injection}
	P_{\mathrm{inj},t}^{\,h}= \tilde{P}_{\mathrm{PV},t}^{\,h}-P_{\mathrm{AC},t}^{\,h}-\tilde{P}_{\mathrm{UL},t}^{\,h}\\
	\label{eq:local-opt-reactive-power-injection}
	Q_{\mathrm{inj},t}^{\,h}=\tilde{Q}_{\mathrm{PV},t}^{\,h}-Q_{\mathrm{AC},t}^{\,h}-\tilde{Q}_{\mathrm{UL},t}^{\,h}\\
	\label{eq:local_thermal_dynamics}
	\begin{gathered}
 T_{\mathrm{AC},t+1}^{\,h}=\exp\left(-\Delta t/R_{\mathrm{AC}}^{\,h}\, C_{\mathrm{AC}}^{\,h}\right)\cdot T_{\mathrm{AC},t}^{\,h}
\\ + \Biggl(1-\exp\left(-\Delta t/R_{\mathrm{AC}}^{\,h}\,  C_{\mathrm{AC}}^{\,h}\right)\Biggr) \left(T_{\mathrm{out},t}^{\,h}-\eta_{\mathrm{AC}}^{\,h}\cdot R_{\mathrm{AC}}^{\,h}\cdot P_{\mathrm{AC},t}^{\,h}\right)
	\end{gathered}\\
	\label{eq:local_thermal_limits}
	 \underline{T}_{\mathrm{AC}}^{\,h} \leq T_{\mathrm{AC},t}^{\,h} \leq \overline{T}_{\mathrm{AC}}^{\,h}\\
	 \label{eq:local_operating_envelopes}
	\mathbf{A}\cdot \begin{bmatrix} P_{\mathrm{inj},t}^{\,h} \\ Q_{\mathrm{inj},t}^{h} \end{bmatrix} \leq \mathbf{b}\\
	\text{and \eqref{eq:active-power-bounds-AC}-\eqref{eq:uncontrollable-load-power-relationship}} \nonumber
\end{gather}
\end{subequations}
where $T_{\mathrm{AC},t}^{\,h}$ is the indoor temperature, $T_{\mathrm{out},t}^{\,h}$ is the outdoor temperature and $\pi_{t}^{h}$ is the electricity price for house $h$ at time $t$. $\underline{T}_{\mathrm{AC}}^{\,h}$ and $\overline{T}_{\mathrm{AC}}^{\,h}$ represent lower and upper limits of thermal comfort, $R_{\mathrm{AC}}^{\,h}$ is the thermal resistance, $C_{\mathrm{AC}}^{\,h}$ is the thermal capacitance for the air-conditioning system and $\Delta t$ is the sampling time \cite{StateEstimationMathieu}. Assuming \textit{DOE customers} are exposed to market prices \cite{WholesaleDR}, maximising financial gains by exporting power to the grid is represented by \eqref{eq:household_obejctive_function}, active and reactive power balance at the POC are represented by \eqref{eq:local-opt-active-power-injection} and \eqref{eq:local-opt-reactive-power-injection}, thermal dynamics of the inverter-type air-conditioner is represented by \eqref{eq:local_thermal_dynamics}, thermal comfort limits are represented by \eqref{eq:local_thermal_limits}. The DOE constraints at the POC is represented by \eqref{eq:local_operating_envelopes}.
\par
The objective of the DRA is to provide DR services by tracking the load set-point signal commanded by the market operator. This is achieved by controlling the consumption of air-conditioners belonging to \textit{DOE customers}. This can be mathematically expressed as:
\begin{equation}
	\label{eq:aggregator_objective}
	\underset{P_{\mathrm{AC},t}^{\,h}}{\mathrm{min}}\quad \left(\sum_{h\in\DOEuser} P_{\mathrm{AC},t}^{\,h}-P_{\mathrm{ref},t}\right)^2
\end{equation}
where $P_{\mathrm{ref},t}$ is the reference load set-point sent by the market operator at time $t$. According to \eqref{eq:aggregator_objective}, the squared deviation of total air-conditioner power consumption from the load set-point is minimised.
\par
The DRA could centrally control air-conditioning loads to track the load set-point signal \cite{Lu_evaluation}. However, it will lead to end-user privacy violations \cite{NIST_privacy}. To overcome this, a hierarchical implementation based on the ADMM form of the \textit{resource sharing problem} \cite{Boyd_ADMM} is utilised. Based on this implementation, a local controller at each \textit{DOE customer} and a coordinating controller at the DRA are established. 

\subsubsection{Local controller problem}
\label{subsubsec: Local controller problem}

Considering household $h\in\DOEuser$, the local controller problem can be formulated as follows:
\begin{gather}
	\label{eq: local controller problem}
	\begin{gathered}
	P_{\mathrm{AC},t}^{\,h^{(\nu+1)}}=\underset{P_{\mathrm{AC},t}^{\,h}}{\mathrm{argmin}}\Bigg(-\pi_{t}^{h}\cdot P_{\mathrm{inj},t}^{\,h}+(\rho/2)\cdot \\
	\left(P_{\mathrm{AC},t}^{\,h}-P_{\mathrm{AC},t}^{\,h^{(\nu)}}+P_{\mathrm{avg},t}^{(\nu)}-P_{t}^{(\nu)}+\theta_{t}^{{(\nu)}}\right)^2\Bigg)
	\end{gathered}\\
\text{subject to \eqref{eq:active-power-bounds-AC}-\eqref{eq:uncontrollable-load-power-relationship}, \eqref{eq:local-opt-active-power-injection}-\eqref{eq:local_operating_envelopes}} \ \nonumber
\end{gather}
where $(\nu)$ represents $\nu$-th iteration of the ADMM scheme, $\rho>0$ is the 
augmented Lagrangian parameter, $P_{\mathrm{avg},t}$ is the average 
of $P_{\mathrm{AC},t}^{\,h}$, $P_{t}$ is an auxiliary variable such that $P_{t}=(1/\vert \DOEuser \vert)\sum_{h\in\DOEuser}P_{t}^{h}$, and $\theta$ is the dual scaled variable. The interested readers are referred to \cite{Boyd_ADMM} for more details on the scaled form of ADMM for the \textit{sharing problem}.

\subsubsection{Coordinating controller problem}
\label{subsubsec: Coordinating controller problem}

For the coordinating controller at the DRA, the load set-point tracking problem in \eqref{eq:aggregator_objective} is modified as:
\begin{subequations}
\begin{gather}
	\label{eq:coordinating-controller-tracking-objetive}
	\begin{gathered}
			P_{t}^{(\nu+1)}=\underset{P_{t}}{\mathrm{argmin}}\Biggl(\Bigl(\lvert\DOEuser\rvert P_{t}-P_{\mathrm{ref},t}\Bigr)^2+\\\Big(\vert \DOEuser \vert \rho/2\Big) \cdot \Bigl(P_{t}-\theta_{t}^{{(\nu)}}-P_{\mathrm{avg},t}^{(\nu+1)} \Bigr)^2 \Biggr)
	\end{gathered}\\
	\label{eq:coordinating-controller-theta-update}
	\theta_{t}^{{(\nu+1)}}=\theta_{t}^{{(\nu)}}+P_{\mathrm{avg},t}^{(\nu+1)}-P_{t}^{(\nu+1)}
\end{gather}
\end{subequations}
The tracking problem is represented by \eqref{eq:coordinating-controller-tracking-objetive} and the global dual variable update is represented by \eqref{eq:coordinating-controller-theta-update}.
\par
To further explain this, in iteration $\nu$ at time step $t$, each local controller solves \eqref{eq: local controller problem} to determine $P_{\mathrm{AC},t}^{\,h^{(\nu+1)}}$ for all $h\in\DOEuser$ and passes to the coordinating controller. In the next step, the coordinating controller calculates the average of $P_{\mathrm{AC},t}^{\,h^{(\nu+1)}}$ which is given by $P_{\mathrm{avg},t}^{(\nu+1)}$ and thereafter, solves \eqref{eq:coordinating-controller-tracking-objetive} to determine $P_{t}^{(\nu+1)}$. Finally, the coordinating controller updates the global dual variable as in \eqref{eq:coordinating-controller-theta-update}. This process is repeated at each iteration until the following termination criteria are met. 
\begin{equation}
	\label{eq:termination criteria}
	\| \mathbf{r}_{t}^{(\nu)} \|_{2} \leq \epsilon^{\textrm{prim}} \quad \text{and} \quad  \| \mathbf{s}_{t}^{(\nu)} \|_{2} \leq \epsilon^{\textrm{dual}}
\end{equation}
where $\mathbf{r}_{t}^{(\nu)}$ and $\mathbf{s}_{t}^{(\nu)}$ are the primal and dual residuals in $\nu$-th iteration at time $t$, $\epsilon^{\textrm{prim}}$ and $\epsilon^{\textrm{dual}}$ are the tolerances for primal and dual residual respectively. Furthermore, $ \mathbf{r}_{t}^{(\nu)}=[P_{\mathrm{AC},t}^{\,1^{(\nu)}}-P_{t}^{(\nu)},\ldots,P_{\mathrm{AC},t}^{\,n^{(\nu)}}-P_{t}^{(\nu)}]$ for $\DOEuser:=\{1,2\ldots,n\}$ and $\mathbf{s}_{t}^{(\nu)}=[P_{t}^{(\nu+1)}-P_{t}^{(\nu)}]$. In addition to \eqref{eq:termination criteria}, termination criteria based on maximum ADMM iteration, \texttt{maxiter}, is also utilised to speed up ADMM convergence and dispatch power consumption set-points for air-conditioning loads in real-time.
\begin{figure}[t]
	\centering
	\includegraphics[width=0.75\columnwidth]{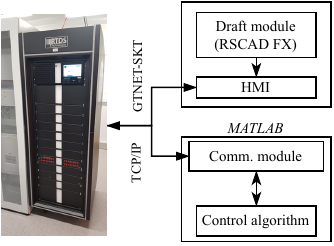}
	\caption{The overall SIL setup}
		\label{fig:SIL setup}
\end{figure}

\section{Software-in-the-loop validation setup}
\label{sec: Software-in-the-loop setup}

\begin{figure*}[t]
	\centering
		\includegraphics[scale=0.35]{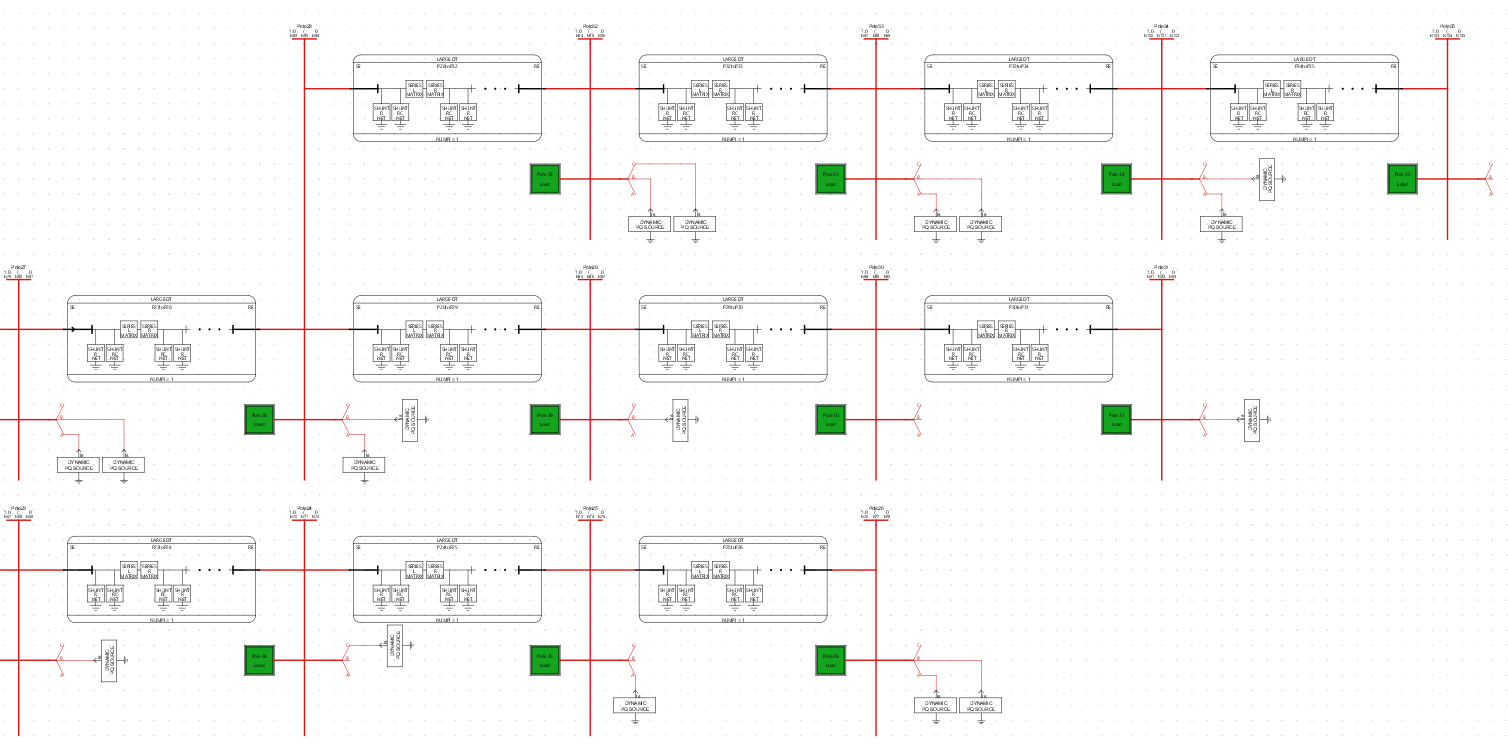}
		\caption{A snapshot of the network modelled in RSCAD FX 1.3.1}
        \label{fig:modelling_of_the_network_RSCADFX}
\end{figure*}

To study the efficacy of the proposed control scheme in a realistic setting, software-in-the-loop (SIL) simulations are performed in the real-time digital simulator (RTDS) platform. Once the control algorithm determines control set points at a certain time step, these set-points are given as inputs to the RTDS and thereafter, grid data is retrieved as outputs from the RTDS with the support of the control processing unit and the digital communication interface. A block diagram of the overall SIL setup for the study is given in Fig. \ref{fig:SIL setup}.
\par
The draft module of network in the case study is modelled in RSCAD FX 1.3.1  \cite{RDSCADFX_quick_start_guide}. Since the network represents a residential distribution feeder, \textit{distribution mode} in RSCAD FX with components from the distribution library are used. To model household loads (including both controllable and uncontrollable load) for all customers, 1-phase dynamic loads of RL type with external P \& Q controls are used. To model rooftop PV generation for \textit{DOE customers} and \textit{Non-DOE customers}, 1-phase dynamic PQ sources of P/Q control type with externally assigned P/Q control set-points are chosen. To model three-phase distribution lines, cascaded PI models (MATPI) with series [R] and [L] matrices specified in the form of lower triangular matrices are used. Furthermore, single-phase RMS meters are placed in the draft module to measure the RMS voltage of selected nodes in real-time. The time step for the simulation in distribution mode is set to 150.0 microseconds. A snapshot of the overall modelling approach in RSCAD FX is shown in Fig.~\ref{fig:modelling_of_the_network_RSCADFX}.
\par
The human machine interface (HMI) is a desktop PC with Intel(R) Core(TM) i7-9700 CPU of 3.00 GHz and 64.0 GB memory. The workstation is installed with RSCAD FX 1.3.1. The workstation is installed with RSCAD FX 1.3.1. Moreover, the control algorithm is written in MATLAB 2018b together with YALMIP toolbox \cite{Lofberg2004} and MPT3 \cite{MPT3_toolbox} in the workstation PC. The RTDS is equipped with a NovaCor(R) processor---each an IBM Power8 processor with 10 cores operating at 3.5 GHz. For the SIL study, 3 processor cores are used. Furthermore, the RTDS is equipped with a GTNETx2 card to interface with the workstation (HMI). 
\par
The communication between MATLAB and the RTDS is established via the TCP/IP protocol whereas GTNET-SKT protocol is used to establish the communication between the RTDS and the HMI as shown in Fig.~\ref{fig:SIL setup}. 
Furthermore, 39 data points are sent to GTNET-SKT and then passed to the HMI. This corresponds to real-time rms voltage measurements of chosen nodes obtained from the grid simulator. The number of data points received from the GTNET-SKT (sent by the HMI) is 296. This corresponds to P, Q set-points for all household loads and P\&Q set-points for households with PV, i.e., \textit{DOE customers} and \textit{Non-DOE customers}, in the case study. While inputs to the grid simulator are updated every 5-mins (aligned with the sampling time of the control algorithm), outputs are obtained every 30 seconds from the grid simulator (clock signal is also modelled in RSCAD FX). Moreover, a MATLAB script is written for initialising the TCP/IP connection between the RTDS and MATLAB, sending inputs to RTDS every 5-mins and obtaining RTDS outputs every 30 seconds.

\section{Results}
\label{sec: Results}
\begin{figure*}[t]
	\centering
	\includegraphics[scale=0.85]{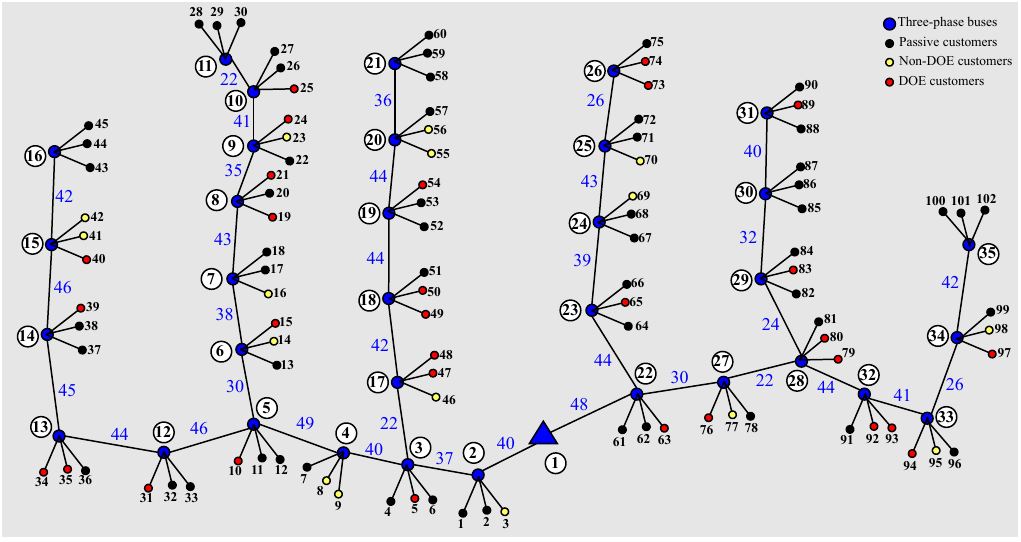}
	\caption{The diagram of the LV residential network \cite{WANG2019113927}; \Circled{1} (blue triangle) represents the LV distribution transformer; blue dots (\Circled{2} -- \Circled{35}) represent three-phase P-Q buses (poles); black lines represent overhead conductors; italicised numbers in blue represent distances in metres}
	\label{fig:LV network}
\end{figure*}

Both computer-based simulation validations and SIL validations are performed on a practical residential network in Queensland, Australia \cite{WANG2019113927}. The network data is only available up to the pole-level and it is assumed that a single customer is connected to each phase of three-phase P-Q buses represented by \Circled{2}-\Circled{35} in  Fig. \ref{fig:LV network}. This results in a total of $34\times 3=102$ single-phase customers. It is assumed that approximately $45\%$, i.e., 46 out of 102 households, are equipped with rooftop PV generation \cite{Australian_PV_institute}. Out of 46 households with rooftop PV generation, 30 corresponds to \textit{DOE customers} and 16 corresponds to \textit{Non-DOE customers} as in Fig. \ref{fig:LV network}. The ratings of the PV inverters are [3.0, 3.6, 4.0, 5.0, 6.0, 8.0] kWp \cite{web_SunnyBoy} with  $\varphi_{{\mathrm{PV}}}^{\,h}=0.8$ for all $h\in \DOEuser\cup \PVuser$ aligned with \cite{AS4777_standards,Energex_standards_for_small_connections}. For uncontrollable loads, $\varphi_{{\mathrm{UL}}}^{\,h}=0.95$ (lagging) for all $h\in\mathcal{H}$ \cite{Energex_standards_for_small_connections}. The household uncontrollable load and rooftop PV profiles are obtained from \cite{CREST_demand_tool}. The market price of electricity $(\pi^{h})$ for all $h$ is obtained from the National Electricity Market (NEM) \cite{NEMdashboard}. Furthermore, import and export limits remain at 10 kW and 5 kW for \textit{passive customers} and \textit{non-DOE customers} respectively \cite{Energex_standards_for_small_connections}.
\par
For \textit{DOE customers}, i.e., $h\in \DOEuser$, the rated power of air-conditioners participating in DR is $\overline{P}_{\mathrm{AC}}^{\,h}\sim \mathcal{N}(2.5, 3.5)$ kW with $\varphi_{{\mathrm{AC}}}^{\,h}=0.95$ (lagging) aligned with \cite{active_power_factor_correction}. The thermal parameters $R_{\mathrm{AC}}^{\,h}\sim \mathcal{N}(1.5, 2.5)$ \textdegree C/kW; $C_{\mathrm{AC}}^{\,h}\sim \mathcal{N}(1.5, 2.5)$ \textdegree kWh/\textdegree C and $\eta_{\mathrm{AC}}^{\,h}=2.5$ \cite{StateEstimationMathieu}. The thermal comfort limits are	$\underline{T}_{\mathrm{AC}}^{\,h}=22$\textdegree C and $\overline{T}_{\mathrm{AC}}^{\,h}=24$\textdegree C. Furthermore, the initial indoor temperature is assumed to be at $23$\textdegree C for all \textit{DOE customers}. The outdoor temperature profile is obtained from \cite{uqweatherdata}. To validate the proposed control scheme, it is assumed that the DRA receives a load set-point signal that should be tracked by air-conditioning loads for a period of 2-hours, starting from 10:00 and ending at 12:00, on 06-12-2020. The load set-point signal is constructed as in \cite{Lu_evaluation} with indoor temperature set-point at 23\textdegree C and a regulation capacity of 20\% from the baseline consumption.
\par
For probabilistic load flow calculations in section~\ref{subsec: DNSP determining operating envelopes in real-time}, the three-phase network in Fig. \ref{fig:LV network} is modelled in OpenDSS \cite{OpenDSS}. The voltage statutory limits are set at $\underline{v}=0.94$ pu and $\overline{v}=1.10$ pu aligned with \cite{Energex_standards_for_small_connections}. The overall control algorithm is written in MATLAB on the HMI. The optimisation problems in section \ref{subsec: DRA controlling household loads in real-time} are modelled with YALMIP toolbox \cite{Lofberg2004} and solved with Gurobi 9.1.2 \cite{gurobi}. The \textit{convex hull} and its half-space representation are calculated using MPT3 \cite{MPT3_toolbox}. For the ADMM iterative algorithm, $\epsilon^{\textrm{r}}$ and $\epsilon^{\textrm{s}}$ are \texttt{1e-3}, $\rho=$1 and the additional termination criteria, \texttt{maxiter} = 15. The simulation step size $\Delta t=$5-min, i.e., DOE assignment by the DNSP and real-time control operation of the DRA occurs every 5-mins throughout the DR period. This is aligned with the market clearing interval of the NEM \cite{dyanmic_envelopes_outcome_report}. For the DNSP to establish operating envelopes for \textit{DOE customers} as discussed in section~\ref{subsec: DNSP determining operating envelopes in real-time}, 500 probabilistic load flow scenarios $(\vert \Omega \vert)$ are considered. 
\par
\subsection{The estimation of DOEs by the DNSP}
\label{subsec:DNSP DOE estimation}
\begin{figure}[t]
	\centering
	\subfloat[Bounding box (capability curve)\label{fig:bounding_box}]{%
		\includegraphics[width=0.50\linewidth]{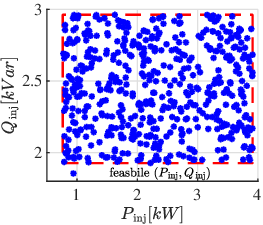}}
	\hfill
	\subfloat[Convex hull\label{fig:convex_hull}]{%
		\includegraphics[width=0.50\linewidth]{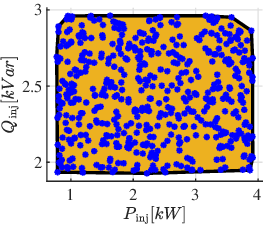}}
	\caption{The bounding box representing the capability curve and the convex hull representing the envelope at the POC for house 38 connected to bus \Circled{14} at 11:15 ($\vert \Omega \vert = 500$).}
	\label{fig:convex_hull_estimation}
\end{figure}

Fig.~\ref{fig:convex_hull_estimation} illustrates the construction of bounding box \eqref{eq:bouding box approaxmation} and the \textit{convex hull} for household 39 (\textit{DOE customer} connected to bus \Circled{14} in Fig.~\ref{fig:convex_hull_estimation}) at 11:15. 
As can be seen from Fig. \ref{fig:bounding_box}, once a household calculates maximum and minimum power injections at the POC, thereafter, the DNSP determines the bounded region of operation (represented by the red dashed region) of the household for that time step. Afterwards, probabilistic load flow studies are performed to determine feasible P-Q injection pairs (highlighted in blue) within the bounded region for which resulting three-phase load flows would not result in voltage violations. Fig. \ref{fig:convex_hull} illustrates the DOE of household 39 obtained by constructing the \textit{convex hull} of all feasible P-Q injection pairs. It can be clearly observed that DOEs capture both active and reactive, import and export power limits at household connection point.
\begin{figure}[t]
	\centering
	\subfloat[Tracking performance of the DRA\label{fig:tracking_performance}]{%
		\includegraphics[width=\columnwidth]{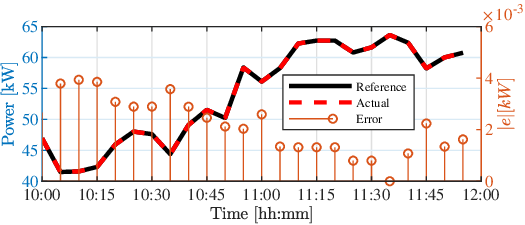}}
	\hfill
        \subfloat[The variation of air-conditioner power for \textit{DOE customers}\label{fig:air_conditioner_power_for_DOE_customers}]{%
		\includegraphics[width=\columnwidth]{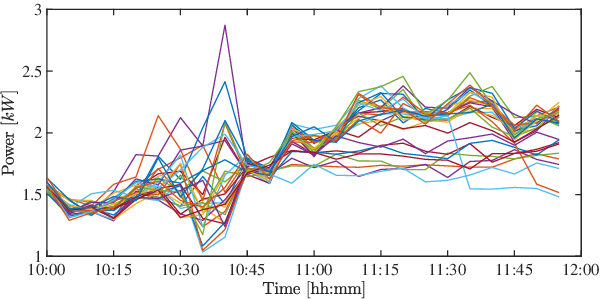}}
        \hfill
	\subfloat[The variation of indoor temperature for \textit{DOE customers}\label{fig:indoor_temperature_profile}]{%
		\includegraphics[width=\columnwidth]{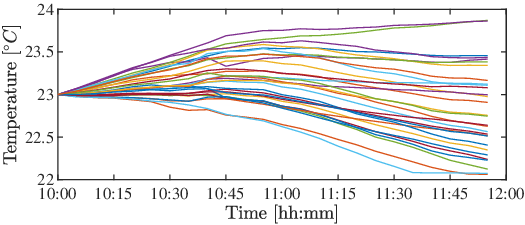}}
	\caption{Tracking performance and the variation of indoor temperature during the DR event}
	\label{fig:overall performance}
\end{figure}
\par
\subsection{Real-time control action of the DRA}
\label{subsec:Real-time control action of the DRA}
The aggregate tracking performance of the DRA in providing DR from air-conditioners in real-time is illustrated in Fig.~\ref{fig:tracking_performance}. According to the figure, the DRA is able to track the load set-point signal from 10:00 to 12:00 with a high level of accuracy (tracking error $< 0.004$ kW) by controlling the consumption of inverter-type air-conditioners belonging to 30 \textit{DOE customers} in the residential feeder. Looking at the variation of air-conditioner power consumption as in Fig.~\ref{fig:air_conditioner_power_for_DOE_customers}, it is observed that the power profile tends to follow an identical to the commanded set-point by the system operator unless household local controllers cannot maintain indoor thermal comfort if identical profile is followed. During these periods, household local controllers' set-points diverge from the identical profile. This is clearly observed in both Fig.~\ref{fig:air_conditioner_power_for_DOE_customers} and Fig.~\ref{fig:indoor_temperature_profile} between 10:15 and 10:45 and also from 11:00 until 12:00. Nonetheless, from Fig.~\ref{fig:indoor_temperature_profile} it can be clearly observed that indoor temperature is maintained within thermal comfort limits  $[22,24]$\textdegree C for all \textit{DOE customers} for the duration of the DR event. Therefore, it can be concluded that the proposed hierarchical implementation for the DRA preserves end-user thermal comfort while providing DR by tracking the load set-point signal.
\begin{figure}[t]
    \centering
    \subfloat[DOE customers\label{fig:DOE_customer_injection}]{%
		\includegraphics[width=\columnwidth]{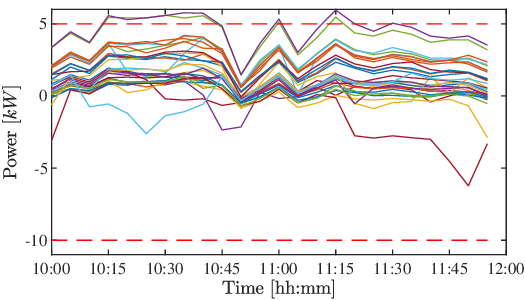}}
        \hfill
        \subfloat[Non-DOE customers\label{fig:Non_DOE_customer_injection}]{%
		\includegraphics[width=\columnwidth]{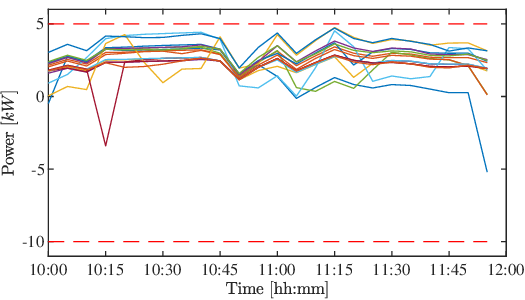}}
        \hfill
        \subfloat[Passive customers\label{fig:passive_customer_injection}]{%
		\includegraphics[width=\columnwidth]{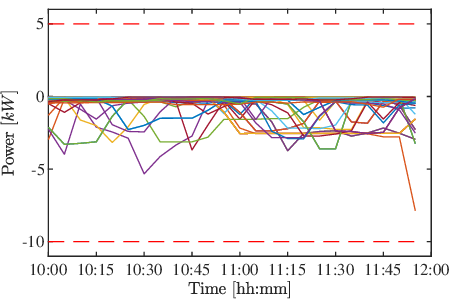}}
  	\caption{The variation of active power injections at the POC of households}
	\label{fig:active_power_injections}
\end{figure}
\par
\begin{figure*}[t]
	\centering	
        \includegraphics[scale=0.75]{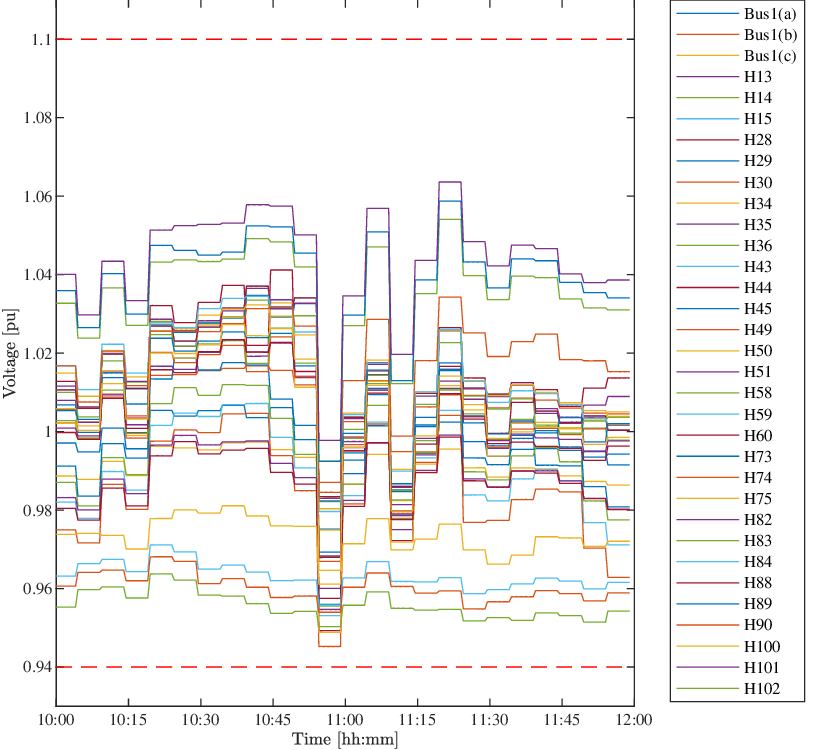}
	\caption{Voltage profile of selected nodes during for the DR period; red dotted lines represent lower and upper voltage limits}
	\label{fig:voltage_profile_RTDS}
\end{figure*}
The corresponding variation of net active power injections at the POC during the DR period for all customers is depicted in Fig.~\ref{fig:active_power_injections}. It can be clearly seen from Fig.~\ref{fig:DOE_customer_injection} that \textit{DOE customers} are able to export active power beyond the 5~kW fixed limit during the DR event. To further analyse this, when the system operator's commanded set-point is below the baseline consumption, DRA dispatches control set-points to lower the consumption of air-conditioners. Since rooftop PV of \textit{DOE customers} are already exporting power to the grid during period, this will lead to an increase in net power exports to the grid at the POC. On the contrary, when commanded set-point is above the baseline consumption, DRA sends set-points to increase the consumption of air-conditioners (from the previous set-point). Consequently, the net power exports at the POC decreases. This is clearly observed between 10:45 and 11:00 where the DRA ramps up tracking, therefore, the net power injection of \textit{DOE customers} reaches a dip (in Fig.~\ref{fig:DOE_customer_injection}). Looking at Fig.~\ref{fig:Non_DOE_customer_injection}, it is apparent that active power exports are capped at 5 kW for \textit{Non-DOE customers}. Although rooftop PV is considered to be uncontrollable for \textit{Non-DOE customers}, in the event of net power injections beyond 5 kW, the DNSP curtails rooftop PV generation. For  \textit{passive customers}, since they are not equipped with rooftop PV generation, net injections remains below zero during the DR event as seen in Fig.~\ref{fig:passive_customer_injection}. Moreover, the fluctuation in the net power injection profile corresponds to the behaviour of uncontrollable loads. On the other hand, it is also noted that for all three customer categories, the DR action from 10:00 to 12:00 will not result in net injections reaching 10 kW fixed import limit (-10 kW with respect to power injections). This can be understood by the fact that household net consumption tends to be negative during the day-period (10:00 to 12:00 in this study) due to the significant contribution of rooftop PV exports compared to load reductions due to DR. However, this is irrelevant for \textit{passive customers} as they do not participate in DR under the DOE framework and do not possess rooftop PV generation.
\par
\subsection{Software-in-the-loop validation}
\label{subsec:SIL validation results}
\vspace{-3pt}
The voltage outputs of bus \Circled{1} and a chosen set of household nodes at each sampling instant obtained from the RTDS SIL study is shown in Fig.~\ref{fig:voltage_profile_RTDS}. Although 5 kW export limits are exceeded for certain \textit{DOE customers}, the resulting voltage profile does not show over-voltage issues. The dip in the voltage profile between 10:45 and 11:00 is due to the ramping up of load set-point resulting in increasing the air-conditioner power set-point of \textit{DOE customers}. This increase in load (or else decrease in net active power exports) leads to lower voltage that causes the dip in the voltage profile. Nonetheless, the voltage profile is maintained within technical limits, $\left[0.94, 1.10\right]$ pu, for the duration of the DR event. Hence, it can be concluded that the DOEs established by DNSP ensures voltage network limits during the real-time control operation of the DRA.
\par
On the other hand, the total execution time for a DR duration of 2-hours (from 10:00 to 12:00) is approximately 1267 seconds ($\approx$ 22-mins). In other words, the average execution time per sampling instant is 52 seconds. This includes probabilistic load flows and determining the overall envelope in \textit{Stage I} and ADMM-based control of air-conditioners to provide DR in \textit{Stage II}. Since the sampling interval ($\Delta t$) is set to be 5-mins to align with the operation of the NEM, it can be concluded that the overall approach is scalable in a practical setting.
\par
\begin{remark}
    For the DR operation, \textit{DOE customers} do not need to share information on $\overline{P}_{\mathrm{AC}}$,  $R_{\mathrm{AC}}$, $C_{\mathrm{AC}}$ and $\eta_{\mathrm{AC}}$ with the DRA. Instead they share household-specific information with the local controller present in their own premises under the ADMM scheme. Hence, the overall approach preserves end-user data privacy.
\end{remark}

% \begin{figure}[b]
%     \centering
%     \subfloat[primal residual\label{ig:primal_residual_APPEEC}]{%
% 		\includegraphics[scale=0.75]{images/primal_residual_APPEEC_work.eps}}
%         \hfill
%         \subfloat[dual residual\label{fig:dual_residual_APPEEC}]{%
% 		\includegraphics[scale=0.75]{images/dual_residual_APPEEC_work.eps}}
%   	\caption{Convergence of the ADMM control scheme for the DRA}
% 	\label{fig:convergence_ADMM_for_DRA}
% \end{figure}
\par
% Fig.~\ref{fig:convergence_ADMM_for_DRA} shows the convergence of the ADMM hierarchical implementation for the DRA. When analysing the results, it is observed that at each sampling step, the primal and dual residuals settle after a couple of iterations (approximately 2 iterations). Thereafter, the dual residual remains at almost zero but the primal residual tends to remain at the settled value without further improving to zero. Hence, the additional termination criteria, \texttt{maxiter}, is used to terminate the convergence at one time instant and move to the next step. Moreover, the value of \texttt{maxiter} is set to be 15 as there is no significant improvement is seen in both profiles from 2nd iteration to 15th iteration. With aforementioned criteria, the total execution time for the two-stage control for a DR duration of 2-hours (from 10:00 to 12:00) is approximately 1267 seconds ($\approx$ 22-mins). In other words, the average execution time per sampling instant is 52 seconds. Since the sampling interval ($\Delta t$) is set to be 5-mins to align with the operation of the NEM, it can be concluded that the overall approach is scalable.

\section{Conclusions}
\label{sec: Conclusions}
In this work, a real-time coordinated control scheme is proposed for residential DR participation under the DOE framework. In the first stage, the DNSP employs a \textit{convex hull} technique to construct DOEs at end-user POCs. In the second stage, the DRA adopts a hierarchical control scheme for providing DR while utilising DOEs shared by the DNSP. Simulation validations suggests that precise DR can be obtained without jeopardising LV network voltage limits while maintaining indoor temperature within comfort limits. Moreover, the DOE approach enables active power exports beyond the 5 kW fixed limits. In addition to that, the overall approach is scalable and compliant with the market dispatch interval of the NEM.

\section*{Acknowledgment}
The authors would like to thank for the support given by the University of Queensland -- Centre for Energy Innovation (UQ-CEDI) under the Advance Queensland grant (grant no: AQPTP01216-17RD1). The authors would also like thank Prof. Jovica Milanovi\'c for providing valuable feedback on the manuscript.

\bibliographystyle{IEEEtran}
\bibliography{references}

\begin{thebibliography}{10}
\providecommand{\url}[1]{#1}
\csname url@samestyle\endcsname
\providecommand{\newblock}{\relax}
\providecommand{\bibinfo}[2]{#2}
\providecommand{\BIBentrySTDinterwordspacing}{\spaceskip=0pt\relax}
\providecommand{\BIBentryALTinterwordstretchfactor}{4}
\providecommand{\BIBentryALTinterwordspacing}{\spaceskip=\fontdimen2\font plus
\BIBentryALTinterwordstretchfactor\fontdimen3\font minus \fontdimen4\font\relax}
\providecommand{\BIBforeignlanguage}[2]{{%
\expandafter\ifx\csname l@#1\endcsname\relax
\typeout{** WARNING: IEEEtran.bst: No hyphenation pattern has been}%
\typeout{** loaded for the language `#1'. Using the pattern for}%
\typeout{** the default language instead.}%
\else
\language=\csname l@#1\endcsname
\fi
#2}}
\providecommand{\BIBdecl}{\relax}
\BIBdecl

\bibitem{Energex_standards_for_small_connections}
{ENERGEX Limited}, ``{STNW3510: Dynamic Standard for Small IES Connections},'' \url{https://www.ergon.com.au/__data/assets/pdf_file/0004/962779/STNW3510-Dynamic-Standard-for-Small-IES-Connections.pdf}.

\bibitem{CellPress_Tushar}
W.~Tushar, M.~I. Azim, M.~R. Alam, C.~Yuen, R.~Sharma, T.~Saha, and H.~V. Poor, ``{Achieving the UN’s sustainable energy targets through dynamic operating limits},'' \emph{iScience}, vol.~26, no.~7, p. 107194, 2023.

\bibitem{dyanmic_envelopes_outcome_report}
{Dynamic Operating Envelopes Working Group}, ``{Outcomes Report},'' \url{https://arena.gov.au/assets/2022/03/dynamic-operating-envelope-working-group-outcomes-report.pdf}.

\bibitem{Blackhall2020}
\BIBentryALTinterwordspacing
L.~Blackhall, ``On the calculation and use of dynamic operating envelopes,'' 2020. [Online]. Available: \url{https://arena.gov.au/assets/2020/09/on-the-calculation-and-use-of-dynamic-operating-envelopes.pdf}
\BIBentrySTDinterwordspacing

\bibitem{Nando_OPF_paper}
M.~Z. Liu, L.~F. Ochoa, and S.~Member, ``{Using OPF - Based Operating Envelopes to Facilitate Residential DER Services},'' \emph{IEEE Transactions on Smart Grid}, no. June, p.~1, 2022.

\bibitem{ensuring_petrou}
K.~Petrou, A.~T. Procopiou, L.~Gutierrez-Lagos, M.~Z. Liu, L.~F. Ochoa, T.~Langstaff, and J.~M. Theunissen, ``Ensuring distribution network integrity using dynamic operating limits for prosumers,'' \emph{IEEE Transactions on Smart Grid}, vol.~12, no.~5, pp. 3877--3888, Sep. 2021.

\bibitem{NIST_privacy}
V.~Pillitteri and T.~Brewer, ``{Guidelines for Smart Grid Cybersecurity},'' 2014.

\bibitem{Attarha_network_secure_envelopes}
A.~Attarha, S.~M.~N. R.A., P.~Scott, and S.~Thiebaux, ``{Network-Secure Envelopes Enabling Reliable DER Bidding in Energy and Reserve Markets},'' \emph{IEEE Transactions on Smart Grid}, p.~1, 2021.

\bibitem{Techno_economic_DOE}
M.~R. Alam, P.~T.~H. Nguyen, L.~Naranpanawe, T.~K. Saha, and G.~Lankeshwara, ``Allocation of dynamic operating envelopes in distribution networks: Technical and equitable perspectives,'' \emph{IEEE Transactions on Sustainable Energy}, pp. 1--13, 2023.

\bibitem{DOE_with_flexibility}
G.~Lankeshwara, R.~Sharma, R.~Yan, T.~K. Saha, and J.~V. Milanović, ``Time-varying operating regions of end-users and feeders in low-voltage distribution networks,'' \emph{IEEE Transactions on Power Systems}, pp. 1--12, 2023.

\bibitem{Rigoni_UCD}
V.~Rigoni, D.~Flynn, and A.~Keane, ``{Coordinating Demand Response Aggregation with LV Network Operational Constraints},'' \emph{IEEE Transactions on Power Systems}, vol.~36, no.~2, pp. 979--990, mar 2021.

\bibitem{WholesaleDR}
\BIBentryALTinterwordspacing
{Australian Energy Market Commission}, ``Wholesale demand response mechanism,'' 2020. [Online]. Available: \url{https://www.aemc.gov.au/rule-changes/wholesale-demand-response-mechanism}
\BIBentrySTDinterwordspacing

\bibitem{EV_stats_Australia}
\BIBentryALTinterwordspacing
{Electric Vehicle Council}, ``State of electric vehicles,'' 2023. [Online]. Available: \url{https://electricvehiclecouncil.com.au/wp-content/uploads/2023/07/State-of-EVs_July-2023_.pdf}
\BIBentrySTDinterwordspacing

\bibitem{Project_EDGE_report}
{Australian Energy Market Operator (AEMO)}, ``{Project EDGE: Final Report},'' \url{https://aemo.com.au/-/media/files/initiatives/der/2023/project-edge-final-report.pdf?la=en}, 2023.

\bibitem{lankeshwara_dynamic_2022}
G.~Lankeshwara and R.~Sharma, ``Dynamic {Operating} {Envelopes}-enabled {Demand} {Response} in {Low}-voltage {Residential} {Networks},'' in \emph{2022 {IEEE} {PES} 14th {Asia}-{Pacific} {Power} and {Energy} {Engineering} {Conference} ({APPEEC})}, 2022, pp. 1--7.

\bibitem{MahdaviPaper}
N.~Mahdavi and J.~H. Braslavsky, ``{Modelling and Control of Ensembles of Variable-Speed Air Conditioning Loads for Demand Response},'' \emph{IEEE Transactions on Smart Grid}, vol.~11, no.~5, pp. 4249--4260, sep 2020.

\bibitem{active_power_factor_correction}
{Lazar Rozenblat}, ``{The basics of active power factor correction},'' \url{https://www.powerfactor.us/active.html}.

\bibitem{AS4777_standards}
{Standards Australia}, ``{Grid connection of energy systems via inverters, Part 2: Inverter requirements},'' \url{https://www.standards.org.au/standards-catalogue/sa-snz/other/el-042/as-slash-nzs--4777-dot-2-colon-2020}.

\bibitem{HEMS}
\BIBentryALTinterwordspacing
{Energex}, ``Home {Energy} {Management} {Systems}.'' [Online]. Available: \url{https://www.energex.com.au/home/control-your-energy/smarter-energy/home-energy-management-systems}
\BIBentrySTDinterwordspacing

\bibitem{de2000computational}
M.~T. De~Berg, M.~Van~Kreveld, M.~Overmars, and O.~Schwarzkopf, \emph{Computational geometry: algorithms and applications}.\hskip 1em plus 0.5em minus 0.4em\relax Springer Science \& Business Media, 2000.

\bibitem{MPT3_toolbox}
M.~Herceg, M.~Kvasnica, C.~Jones, and M.~Morari, ``{Multi-Parametric Toolbox 3.0},'' in \emph{Proc.~of the European Control Conference}, Z\"urich, Switzerland, July 17--19 2013, pp. 502--510, \url{http://control.ee.ethz.ch/~mpt}.

\bibitem{StateEstimationMathieu}
J.~L. Mathieu, S.~Koch, and D.~S. Callaway, ``{State Estimation and Control of Electric Loads to Manage Real-Time Energy Imbalance},'' \emph{IEEE Transactions on Power Systems}, vol.~28, no.~1, pp. 430--440, feb 2013.

\bibitem{Lu_evaluation}
N.~Lu, ``{An evaluation of the HVAC load potential for providing load balancing service},'' \emph{IEEE Transactions on Smart Grid}, vol.~3, no.~3, pp. 1263--1270, 2012.

\bibitem{Boyd_ADMM}
S.~Boyd, N.~Parikh, E.~Chu, B.~Peleato, and J.~Eckstein, ``{Distributed Optimization and Statistical Learning via the Alternating Direction Method of Multipliers},'' \emph{Found. Trends Mach. Learn.}, vol.~3, no.~1, pp. 1--122, jan 2011.

\bibitem{RDSCADFX_quick_start_guide}
{RTDS Technologies}, ``{RSCAD FX: Quick Start Guide}.''

\bibitem{Lofberg2004}
J.~L{\"{o}}fberg, ``{YALMIP: A toolbox for modeling and optimization in MATLAB},'' in \emph{Proceedings of the IEEE International Symposium on Computer-Aided Control System Design}, Taipei, Taiwan, 2004, pp. 284--289.

\bibitem{WANG2019113927}
L.~Wang, R.~Yan, and T.~K. Saha, ``{Voltage regulation challenges with unbalanced PV integration in low voltage distribution systems and the corresponding solution},'' \emph{Applied Energy}, vol. 256, p. 113927, 2019.

\bibitem{Australian_PV_institute}
{Australian PV Institute}, ``{Mapping Australian Photovoltaic Installations},'' \url{https://pv-map.apvi.org.au/historical}.

\bibitem{web_SunnyBoy}
{SMA Solar Technology AG}, ``{SUNNY BOY 3.0/3.6/4.0/5.0},'' \url{https://www.sma.de/fileadmin/content/global/specials/documents/falcon-installer/SB30-50-DEN1708-V22web.pdf}.

\bibitem{CREST_demand_tool}
E.~McKenna and M.~Thomson, ``{High-resolution stochastic integrated thermal–electrical domestic demand model},'' \emph{Applied Energy}, vol. 165, pp. 445--461, 2016.

\bibitem{NEMdashboard}
{Australian Energy Market Operator}, ``{NEM data dashboard},'' \url{https://aemo.com.au/en/energy-systems/electricity/national-electricity-market-nem/data-nem/data-dashboard-nem}.

\bibitem{uqweatherdata}
{School of Earth and Environmental Sciences, The University of Queensland}, ``{UQ weatherstations},'' \url{http://ww2.sees.uq.edu.au/uqweather/}.

\bibitem{OpenDSS}
R.~C. Dugan and T.~E. McDermott, ``{An open source platform for collaborating on smart grid research},'' in \emph{2011 IEEE Power and Energy Society General Meeting}, 2011, pp. 1--7.

\bibitem{gurobi}
{Gurobi Optimization, L L C}, ``{Gurobi Optimizer Reference Manual},'' \url{http://www.gurobi.com}, 2021.

\end{thebibliography}

\end{document}